\documentclass[10pt]{article}
\usepackage[colorlinks=true,citecolor=blue,linkcolor=blue]{hyperref}
\usepackage{multirow}
\usepackage[utf8]{inputenc}
\usepackage[left=2cm,right=2cm,top=2cm,bottom=2cm]{geometry}
\usepackage{cite}
\usepackage{psfrag}
\usepackage[english]{babel}
\usepackage{graphicx}
\usepackage{booktabs}
\usepackage{graphics}
\usepackage{epsfig}
\usepackage{amsmath,amsfonts,amssymb}
\usepackage{pstcol,pst-fill,pst-grad}
\usepackage{euscript}
\usepackage{pstricks}
\usepackage{wrapfig}
\usepackage{slashed}
\usepackage[toc,page]{appendix}
\usepackage{here}
\usepackage{siunitx}
\date{}
\begin{document}
	\title{\vspace{-3cm}
		\hfill\parbox{4cm}{\normalsize \emph{}}\\
		\vspace{1cm}
		{Laser-assisted production of the light charged Higgs boson from top quark  decay in the type-I two Higgs doublet model}}
	\vspace{2cm}
  \author{M. Jakha$^{1,}$\footnote{Corresponding author, E-mail: mohamed.jakha@usms.ma}, S. Mouslih$^{1,2}$, M. Ouhammou$^{1,3}$, R. Chahri$^{1}$, S. El Asri$^{1}$, S. Taj$^{1}$, B. Manaut$^{1}$\\
		{\it {\small$^1$ Laboratoire de Recherche Pluridisciplinaire en Physique (L.R.P.P), }}\\
		{\it {\small Polydisciplinary Faculty, Sultan Moulay Slimane University, Beni Mellal, 23000, Morocco.}}
		\\		
	{\it {\small$^2$ Laboratoire de Physique des Matériaux et Subatomique, Faculty of Sciences, University Ibn Tofail, Kenitra, 14000, Morocco.}}
\\		
	{\it {\small$^3$ Ecole Supérieure de l'Education et de la Formation, Sultan Moulay Slimane University, Beni Mellal, 23000, Morocco.}}	
	}
\maketitle \setcounter{page}{1}
\begin{abstract}
We investigate the impact of a circularly polarized laser field on the top quark decay process into a charged Higgs boson ($t\rightarrow bH^+$) within the type-I two Higgs doublet model. Our study aims to explore how an external electromagnetic field can modify key observables and potentially facilitate the experimental detection of the charged Higgs boson, addressing challenges related to missing energy in collider experiments such as the LHC. Employing the Dirac-Volkov formalism, we model the interaction between charged particles and the laser field and demonstrate that the presence of the laser can notably influence the decay branching ratios under suitable conditions. The analysis reveals that both the intensity and frequency of the laser field play a crucial role in determining the decay width. In particular, for a laser field strength of $3.8\times 10^{14}$ V/cm and a photon energy of $0.117$ eV, the branching ratio of the top quark decaying into a charged Higgs boson with mass in the range $80$-$150$ GeV and a bottom quark reaches $0.97$, surpassing the standard $t\rightarrow bW^+$ channel. These results suggest that strong electromagnetic fields can serve as an effective mechanism to enhance signals of new particles, offering promising avenues for experimental searches beyond the Standard Model.
\end{abstract}

\section{Introduction}
The observation of a Higgs boson with a mass near 125 GeV by the ATLAS \cite{ATLAS} 
and CMS \cite{CMS0} collaborations at the Large Hadron Collider (LHC) represents 
a pivotal achievement in particle physics. This finding effectively completes the 
particle content predicted by the Standard Model (SM), which provides a comprehensive 
framework for describing elementary particles and their interactions. Despite its remarkable achievements, the SM is not a complete theory, 
as it leaves several fundamental questions unresolved, including the nature of dark matter 
and the origin of neutrino masses. This has motivated investigations into extensions 
beyond the SM, among which the Two-Higgs-Doublet Model (2HDM) provides a minimal yet 
attractive framework \cite{2HDM_review}. In the 2HDM, the Higgs sector is expanded by 
introducing a second scalar doublet, resulting in five physical Higgs states: two CP-even 
neutral bosons ($h$, $H$), one CP-odd neutral boson ($A$), and a charged Higgs pair ($H^\pm$).  The discovery of a charged Higgs would thus provide a clear signal of physics beyond the SM. Experimental searches at the LHC have primarily targeted the production and decay  of charged Higgs bosons through their interactions with SM fermions \cite{Higgs_LHC1,Higgs_LHC2}. Nevertheless, detecting such particles remains a considerable challenge, motivating the development of new theoretical approaches to investigate their production mechanisms. Within the 2HDM, the fermion–Higgs couplings define the model's classification. In the type-I 2HDM, all fermions couple exclusively to a single Higgs doublet, while the second doublet does not interact with the fermionic sector. This configuration results in distinctive phenomenological features compared to other 2HDM types, especially concerning Higgs boson production and decay. The type-I framework is widely explored, as it provides a minimal extension of the SM while remaining theoretically consistent and experimentally viable \cite{jueid}.

Simultaneously, laser technology has undergone remarkable progress since its inception in the 1960s. Contemporary high-power laser systems can achieve intensities as high as $10^{22}$ W/cm$^2$ \cite{laser1,laser2}, with even greater intensities anticipated in upcoming facilities. Such advancements have stimulated growing interest in laser-assisted high-energy processes, where strong electromagnetic fields can substantially influence particle interactions and modify the corresponding observables. Ultra-intense laser fields have been proposed as a means to actively detect axions through their excitation \cite{epl3}. Investigations within strong-field quantum electrodynamics (QED) \cite{QED0,QED1,QED2,QED3,QED4,QED5,QED6} and electroweak theory \cite{EW1,epl1,EW2,EW3,EW4,EW5,EW6,EW7,EW8,EW9} have demonstrated that high-intensity laser fields can significantly affect fundamental particle processes, encompassing scattering, decay, and production phenomena. A review presenting the main non-perturbative methods used in the theoretical study of atomic multiphoton processes is given in \cite{epl2}. More specifically, a recent study \cite{EW10} investigated the decay of charged Higgs bosons in the type-II 2HDM under intense circularly polarized laser field, showing that multi-photon processes can significantly alter decay widths and branching ratios. Another recent study \cite{EW11}, conducted within the SM framework, investigated the laser-assisted two-body decay process \( t \to q W^+ \) (with \( q = d, s, b \)) in the presence of a circularly polarized electromagnetic field, showing that laser parameters can significantly modify the decay width, lifetime, and branching ratios of the top quark. The top quark, the heaviest known elementary particle with a mass of $m_t = 172.95 \pm 0.53$ GeV \cite{atlastopmass}, plays a crucial role in this context. Due to its large mass and short lifetime ($\tau_t \simeq 5 \times 10^{-25}$ s), it decays before hadronizing, providing a unique probe for studying new physics. In particular, the decay $t \to bH^+$ in the 2HDM is of special interest, as it could reveal new Higgs sector dynamics \cite{Aaboud2018,Sirunyan2019,Abbaspour}. Understanding how an intense laser field affects this decay process is essential for assessing the feasibility of charged Higgs detection in future collider experiments. High-energy physics experiments demand increasingly powerful particle collisions. With the rapid advancement of modern laser technology, laser acceleration has become an attractive and promising method to enhance collision energies. This makes the study of Higgs production and decay under strong laser fields highly relevant. We refer to studies that have focused on the production of both neutral and charged Higgs bosons in lepton collisions enhanced by laser fields \cite{HLaser1,HLaser2,HLaser3,HLaser4,HLaser5,HLaser6,HLaser7,HLaser8,HLaser9}.

This work explores how a circularly polarized laser field influences the decay of the top quark into a charged Higgs boson within the context of the type-I 2HDM. Using the Dirac-Volkov approach, we evaluate the modifications induced by the laser on both the decay width and the branching ratios. The structure of the paper is as follows: Sec.~\ref{sect.1} introduces the theoretical framework and presents derivations for the decay widths with and without the laser field; Sec.~\ref{sect.2} provides a detailed discussion of the numerical results; and Sec.~\ref{sect.3} offers concluding remarks. Throughout this study, natural units $c = \hbar = 1$ are employed.

\section{Analytical framework for top quark decay}\label{sect.1}
\begin{figure}[hbtp]
\centering
\includegraphics[scale=0.42]{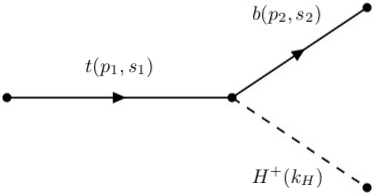}
\caption{Tree-level Feynman diagram for the decay of the top quark into a charged Higgs boson and a bottom quark ($t \rightarrow b H^+$).}\label{diagram}
\end{figure}

This section outlines the theoretical framework underlying the decay process and provides a comprehensive derivation of the decay width, considering both the laser-free case and the modifications induced by the laser field. The corresponding tree-level Feynman diagram is depicted in Fig.~\ref{diagram}.
\\
In the type-I 2HDM, the interaction vertex between the charged Higgs boson 
$H^+$ and the top and bottom quarks is given by
\begin{equation}
\text{vertex}~ t-b-H^+=\frac{\mathrm{g}V_{tb}}{2\sqrt{2}M_W}(\mathcal{A}+\mathcal{B}\gamma^5),
\end{equation}
where the coefficients are defined as $\mathcal{A} = (m_b + m_t) \cot\beta$ and $\mathcal{B} = (m_b - m_t) \cot\beta$. Here, $V_{tb}$ denotes the relevant CKM matrix element, and $\mathrm{g}$ is the electroweak coupling constant. The quantities $m_t$, $m_b$, and $M_W$ correspond to the masses of the top quark, bottom quark, and $W$ boson, respectively.
\subsection{Standard decay $t\rightarrow bH^+$ without laser interaction}
We begin by computing the decay width of the process $t \rightarrow b H^+$ in the absence of a laser field. This subsection serves to provide a clear baseline for comparison and to facilitate a detailed analysis of laser-induced effects.\\
The corresponding tree-level $S$-matrix element in the type-I 2HDM framework is expressed as
\begin{equation}
\begin{split}
S_{fi}(t\rightarrow bH^+)=\frac{-\mathrm{i}\mathrm{g}V_{tb}}{2\sqrt{2}M_W}\int \mathrm{d}^4x\bar{\psi}_b(p_2,s_2)(\mathcal{A}+\mathcal{B}\gamma^5)\psi_t(p_1,s_1)\phi_{H^+}(k_H).
\end{split}
\end{equation}
The various wave functions describing the particles involved in the initial and final states are given by:
 \begin{equation}\label{wave_functions}
\begin{split}
\psi_t(p_1,s_1)&=\frac{u(p_1,s_1)}{\sqrt{2p_1^0V}}\mathrm{e}^{-\mathrm{i}(p_1\cdot x)},\\
\psi_b(p_2,s_2)&=\frac{u(p_2,s_2)}{\sqrt{2p_2^0V}}\mathrm{e}^{-\mathrm{i}(p_2\cdot x)},\\
\phi_{H^+}(k_H)&=\frac{1}{\sqrt{2k_H^0 V}}\mathrm{e}^{\mathrm{i}(k_H\cdot x)},
\end{split}
\end{equation}
where $p_1^0$, $p_2^0$ and $k_H^0$ are, respectively, the temporal components of $p_1$, $p_2$ and $k_H$. Here, $V$ denotes the normalization volume associated with the plane-wave normalization of single-particle states (i.e.\ the wave function is normalized to unity within a box of volume $V$, see Ref.~\cite{greinerQED}); it is a standard quantization parameter in relativistic QED and cancels out in the calculation of physical observables such as decay rates.
\\
Substituting these wave functions and integrating over $\mathrm{d}^4x$, we obtain
\begin{equation}
S_{fi}(t\rightarrow bH^+)=\frac{-\mathrm{i}\mathrm{g}V_{tb}}{2\sqrt{2}M_W}\frac{(2\pi)^4\delta^4(p_2+k_H-p_1)}{\sqrt{8p_1^0p_2^0k_H^0V^3}}\mathcal{M}_{fi},
\end{equation}
where
\begin{equation}
\mathcal{M}_{fi}=\bar{u}(p_2,s_2)(\mathcal{A}+\mathcal{B}\gamma^5)u(p_1,s_1).
\end{equation}
To evaluate the decay width, one considers the squared $S$-matrix element, weighted by the density of available final states. This quantity is then normalized by the total interaction time $T$. In addition, an average over the spin configurations of the initial particles is taken, together with a sum over the spin states of the final particles. In this way, the expression for the decay width is obtained:
\begin{equation}
\begin{split}
\Gamma(t\rightarrow bH^+)=V\int \frac{\mathrm{d}^3p_2}{(2\pi)^3}~V\int \frac{\mathrm{d}^3k_H}{(2\pi)^3}\times |S_{fi}|^2.
\end{split}
\end{equation}
Working in the laboratory frame and following standard techniques in QED, we get the following final expression for the decay width in the absence of the laser field
\begin{equation}
\begin{split}
\Gamma(t\rightarrow bH^+)=\frac{\mathrm{G}_\mathrm{F} \sqrt{2} V_{tb}^2}{16 \pi  m_t^2}\times |\mathbf{p}_2| |\overline{\mathcal{M}_{fi}}|^2,
\end{split}
\end{equation} 
where $\mathrm{G}_\mathrm{F}=\sqrt{2}\mathrm{g}^2/(8M_W^2)$ is the Fermi coupling constant, and
\begin{equation}
\begin{split}
|\overline{\mathcal{M}_{fi}}|^2=\frac{1}{2}\sum_{s_1,s_2} |\mathcal{M}_{fi}|^2=\frac{1}{2}\text{Tr}[(\slashed{p}_2+m_b)(\mathcal{A}+\mathcal{B}\gamma^5)(\slashed{p}_1+m_t)(\mathcal{A}-\mathcal{B}\gamma^5)].
\end{split}
\end{equation}
\subsection{Decay $t\rightarrow bH^+$ under a circularly polarized laser field}
Next, we analyze the decay process under the influence of a circularly polarized laser field, represented by the following four-potential:
\begin{align}\label{potential}
A^{\mu}(\phi)=a^{\mu}_{1}\cos(\phi)+a^{\mu}_{2}\sin(\phi).
\end{align}
In this framework, the laser phase is defined as $\phi = (k \cdot x)$, where $k = (\omega, \mathbf{k})$ denotes the wave four-vector and $\omega$ is the laser frequency. 
The polarization four-amplitudes are given by $a^{\mu}_{1} = |\mathbf{a}|(0,1,0,0)$ and 
$a^{\mu}_{2} = |\mathbf{a}|(0,0,1,0)$, which are orthogonal and have the same magnitude. Consequently, $(a_{1} \cdot a_{2}) = 0$ and $a_{1}^{2} = a_{2}^{2} = a^{2} = -|\mathbf{a}|^{2} = -(\xi_{0}/\omega)^{2}$, where $\xi_{0}$ represents the electric field strength. The four-potential is assumed to satisfy the Lorentz gauge condition, $k_{\mu}A^{\mu}=0$, 
which requires $(k \cdot a_{1}) = (k \cdot a_{2}) = 0$. This condition implies that the propagation vector $\mathbf{k}$ is oriented along the $z$-axis. \\
Now, in the presence of a circularly polarized laser field, the matrix element $S_{fi}$ is written as
\begin{equation}\label{sfic}
\begin{split}
S_{fi}(t\rightarrow bH^+)=\frac{-\mathrm{i}\mathrm{g}V_{tb}}{2\sqrt{2}M_W}\int \mathrm{d}^4x\bar{\psi}_b(p_2,s_2)(\mathcal{A}+\mathcal{B}\gamma^5)\psi_t(p_1,s_1)\phi_{H^+}(k_H).
\end{split}
\end{equation}
The charged Higgs in the final state is not dressed by the field, so its wave function remains the same as that given in equation~(\ref{wave_functions}). In the present analysis, the laser dressing is restricted to the charged fermions, while the charged Higgs boson is treated as an undressed particle. This approximation is adopted to simplify the calculation and to focus on the laser-induced modifications of the production dynamics; a fully dressed treatment of the charged Higgs would lead to additional effective mass shifts and a strong kinematical suppression of the decay. The wave function of the top quark in the presence of the field is given by \cite{volkov}:
\begin{equation}\label{volkov_top}
\psi_t(p_1,s_1)=\bigg[1+\frac{\eta e \slashed{k}\slashed{A}}{2(k\cdot p_1)}\bigg]\frac{u(p_1,s_1)}{\sqrt{2Q_1V}}\times \mathrm{e}^{\mathrm{i}S(q_1,x)}.
\end{equation}
Here, the factor $\eta = -2/3$ reflects the fractional electric charge of the top quark. 
Regarding the sign, we recall that the electron charge is negative, $e = -|e| < 0$. 
The corresponding phase function is then given by
\begin{equation}
S(q_1,x)=-(q_1\cdot x)-\frac{\eta e (a_1\cdot p_1)}{(k\cdot p_1)}\sin(\phi)+\frac{\eta e (a_2\cdot p_1)}{(k\cdot p_1)}\cos(\phi).
\end{equation}
The quantity 
\begin{equation}
q_1 = p_1 - \frac{(\eta e)^2 a^2}{2 (k \cdot p_1)} k,
\end{equation}
corresponds to the quasi-momentum acquired by the top quark in the presence of the laser field, satisfying $q_1^2 = m_t^{*2}$, where $m_t^* = \sqrt{m_t^2 - (\eta e)^2 a^2}$ denotes the effective mass of the top quark within the laser environment.\\
A similar treatment applies to the bottom quark, whose wave function is expressed as \cite{volkov}:
\begin{equation}\label{volkov_bottom}
\psi_b(p_2,s_2)=\bigg[1+\frac{\eta' e \slashed{k}\slashed{A}}{2(k\cdot p_2)}\bigg]\frac{u(p_2,s_2)}{\sqrt{2Q_2V}}\times \mathrm{e}^{\mathrm{i}S(q_2,x)},
\end{equation}
where $\eta'=1/3$, and
\begin{equation}
S(q_2,x)=-(q_2\cdot x)-\frac{\eta' e (a_1\cdot p_2)}{(k\cdot p_2)}\sin(\phi)+\frac{\eta' e (a_2\cdot p_2)}{(k\cdot p_2)}\cos(\phi),
\end{equation}
with
\begin{equation}
q_2=p_2-\frac{(\eta' e)^2a^2}{2(k\cdot p_2)}k~~\text{and}~~m_b^*=\sqrt{m_b^2-(\eta' e)^2a^2}.
\end{equation}
The quantities $Q_1$ and $Q_2$ in equations (\ref{volkov_top}) and (\ref{volkov_bottom}) are the total energies of the top and bottom quarks inside the laser field, such that
\begin{equation}
Q_1=p_1^0-\frac{(\eta e)^2a^2\omega}{2(k\cdot p_1)},~~~~Q_2=p_2^0-\frac{(\eta' e)^2a^2\omega}{2(k\cdot p_2)}.
\end{equation}
Replacing all these wave functions in equation~(\ref{sfic}), we obtain
\begin{equation}\label{sfic1}
\begin{split}
S_{fi}(t\rightarrow bH^+)=\frac{-\mathrm{i}\mathrm{g}V_{tb}}{2\sqrt{2}M_W}\frac{1}{\sqrt{8k_H^0 Q_1 Q_2 V^3}} \int \mathrm{d}^4x \mathrm{e}^{\mathrm{i}(S(q_1,x)-S(q_2,x))} \mathrm{e}^{\mathrm{i}(k_H\cdot x)} \times \mathcal{M}_{fi},
\end{split}
\end{equation}
where the matrix element reads 
\begin{equation}
\mathcal{M}_{fi}=\bar{u}(p_2,s_2)\bigg[1+\frac{\eta' e \slashed{A}\slashed{k}}{2(k\cdot p_2)}\bigg](\mathcal{A}+\mathcal{B}\gamma^5)\bigg[1+\frac{\eta e \slashed{k}\slashed{A}}{2(k\cdot p_1)}\bigg]u(p_1,s_1).
\end{equation}
The first exponential term in Eq.~(\ref{sfic1}) can be recast as
\begin{equation}
S(q_1,x)-S(q_2,x)=(q_2-q_1)\cdot x-z\sin(\phi-\phi_0),
\end{equation}
where
\begin{equation}\label{argument}
\begin{split}
\phi_0&=\arctan(\alpha_2/\alpha_1)~~\text{and}~~z=\sqrt{\alpha_1^2+\alpha_2^2},
\end{split}
\end{equation}
with
\begin{equation}
\begin{split}
\alpha_1&=\frac{\eta e (a_1\cdot p_1)}{(k\cdot p_1)}-\frac{\eta' e (a_1\cdot p_2)}{(k\cdot p_2)},~~\alpha_2=\frac{\eta e (a_2\cdot p_1)}{(k\cdot p_1)}-\frac{\eta' e (a_2\cdot p_2)}{(k\cdot p_2)}.
\end{split}
\end{equation}
Then, we get
\begin{equation}\label{sfi1}
\begin{split}
S_{fi}(t\rightarrow bH^+)=&\frac{-\mathrm{i}\mathrm{g}V_{tb}}{2\sqrt{2}M_W}\frac{1}{\sqrt{8k_H^0 Q_1 Q_2 V^3}} \int \mathrm{d}^4x \bar{u}(p_2,s_2) \Big(C_0+C_1\cos(\phi)+C_2\sin(\phi)\Big)u(p_1,s_1)\\& \times \mathrm{e}^{-\mathrm{i}z\sin(\phi-\phi_0)} \mathrm{e}^{\mathrm{i}(q_2+k_H-q_1)\cdot x},
\end{split}
\end{equation}
where we define 
\begin{equation}
\begin{split}
C_0=&\mathcal{A}+\mathcal{B}\gamma_5,\\
C_1=&C_{p_1}(\mathcal{A}+\mathcal{B}\gamma_5)\slashed{k}\slashed{a}_1+C_{p_2}\slashed{a}_1\slashed{k}(\mathcal{A}+\mathcal{B}\gamma_5),\\
C_2=&C_{p_1}(\mathcal{A}+\mathcal{B}\gamma_5)\slashed{k}\slashed{a}_2+C_{p_2}\slashed{a}_2\slashed{k}(\mathcal{A}+\mathcal{B}\gamma_5),
\end{split}
\end{equation}
with 
\begin{equation}
C_{p_1}=\frac{\eta e}{2(k\cdot p_1)}~~ \text{and}~~ C_{p_2}=\frac{\eta' e}{2(k\cdot p_2)}.
\end{equation}
The three quantities appearing in Eq.~(\ref{sfi1}) can be expanded using the standard identities of ordinary Bessel functions $J_s(z)$ as \cite{landau}
\begin{align}\label{transformation}
\begin{pmatrix}
1\\[1mm]
\cos\phi\\[1mm]
\sin\phi
\end{pmatrix} 
\mathrm{e}^{-\mathrm{i} z \sin(\phi - \phi_0)} 
= \sum_{s=-\infty}^{+\infty} 
\begin{pmatrix}
B_s(z)\\[1mm]
B_{1s}(z)\\[1mm]
B_{2s}(z)
\end{pmatrix} \mathrm{e}^{-\mathrm{i} s \phi},
\end{align}
where
\begin{align}
\begin{pmatrix}
B_s(z)\\[1mm]
B_{1s}(z)\\[1mm]
B_{2s}(z)
\end{pmatrix}
=
\begin{pmatrix}
J_s(z) \mathrm{e}^{\mathrm{i} s \phi_0} \\[1mm]
\dfrac{J_{s+1}(z) \mathrm{e}^{\mathrm{i} (s+1) \phi_0} + J_{s-1}(z) \mathrm{e}^{\mathrm{i} (s-1) \phi_0}}{2} \\[1mm]
\dfrac{J_{s+1}(z) \mathrm{e}^{\mathrm{i} (s+1) \phi_0} - J_{s-1}(z) \mathrm{e}^{\mathrm{i} (s-1) \phi_0}}{2 \mathrm{i}}
\end{pmatrix},
\end{align}
with $z$ being the argument of the Bessel functions defined in Eq.~(\ref{argument}), and $s$ denoting the number of photons exchanged. Equation~(\ref{transformation}) is introduced to expand the laser-induced phase factor into a Fourier series of Bessel functions, which enables the space-time integration in Eq.~(\ref{sfi1}) to be performed analytically. This procedure naturally leads to a summation over the exchanged photon number $s$ and produces the Dirac delta function ensuring energy-momentum conservation, including the contribution of the laser photons $sk$.\\ Applying these expansions to Eq.~(\ref{sfi1}) and integrating over $\mathrm{d}^4 x$, the matrix element $S_{fi}$ takes the form
\begin{equation}
S_{fi}(t\rightarrow bH^+)=\frac{-\mathrm{i}\mathrm{g}V_{tb}}{2\sqrt{2}M_W\sqrt{8k_H^0 Q_1 Q_2 V^3}}\sum_{s=-\infty}^{\infty}\mathcal{M}^{s}_{fi}(2\pi)^{4}\delta^{4}(q_2+k_H-q_1-sk),
\end{equation}
where the quantity $\mathcal{M}^{s}_{fi}$ is defined by
\begin{align}
\mathcal{M}^{s}_{fi}=\bar{u}(p_{2},s_{2})\Big(C_0B_{s}(z)+C_1B_{1s}(z)+C_2B_{2s}(z)\Big)u(p_{1},s_{1}).
\end{align}
Next, the decay width is obtained by multiplying the squared 
$S$-matrix element with the density of final states, performing 
the average over the initial spins, summing over the final spins, 
and finally dividing by the time $T$. This yields
\begin{align}\label{summed}
\Gamma(t\rightarrow bH^+)=\sum_{s=-\infty}^{+\infty}\Gamma^s(t\rightarrow bH^+),
\end{align}
where the decay width $\Gamma^s$ is expressed, for each number of photons $s$, by
\begin{equation}
\Gamma^s(t\rightarrow bH^+) =\frac{N_c \mathrm{G}_\mathrm{F}\sqrt{2}V_{tb}^2}{16(2\pi)^2Q_1}\int \frac{\mathrm{d}^3q_2}{Q_2} \int \frac{\mathrm{d}^3k_H}{k_H^0} \delta^{4}(q_2+k_H-q_1-sk)  |\overline{\mathcal{M}^{s}_{fi}}|^{2},
\end{equation}
where $N_c$ represents the number of colors, taking the value 
$N_c = 3$ for a quark–antiquark pair and $N_c = 1$ in all other cases, and
\begin{align}
|\overline{\mathcal{M}^{s}_{fi}}|^{2}=\frac{1}{2}\sum_{s_{1},s_2}|\mathcal{M}^{s}_{fi}|^{2}.
\end{align}
By performing the integration over $\mathrm{d}^3k_H$ and using the relation $\mathrm{d}^3q_2=|\mathbf{q}_2|Q_2dQ_2d\Omega$, we get
\begin{equation}
\Gamma^s(t\rightarrow bH^+) =\frac{N_c \mathrm{G}_\mathrm{F}\sqrt{2}V_{tb}^2}{16(2\pi)^2Q_1}\int \frac{|\mathbf{q}_2|dQ_2}{k_H^0} \delta^{0}(Q_2+k_H^0-Q_1-s\omega)  |\overline{\mathcal{M}^{s}_{fi}}|^{2}d\Omega,
\end{equation}
with $\mathbf{q}_{2}+\mathbf{k}_{H}-\mathbf{q}_1-s\mathbf{k}=\mathbf{0}$. 
In the top quark rest frame, the remaining integral over $dQ_2$ can be solved by using the familiar formula \cite{greiner}
\begin{align}\label{familiarformula}
\int dxf(x)\delta(g(x))=\dfrac{f(x)}{|g'(x)|}\bigg|_{g(x)=0}.
\end{align}
Thus, we get
\begin{equation}
\Gamma^s(t\rightarrow bH^+) =\frac{N_c \mathrm{G}_\mathrm{F}\sqrt{2}V_{tb}^2}{16(2\pi)^2Q_1}\int \frac{|\mathbf{q}_2|}{k_H^0 |g'(Q_2)|}\times |\overline{\mathcal{M}^{s}_{fi}}|^{2}d\Omega,
\end{equation}
where 
\begin{equation}
g'(Q_2)=1+\frac{Q_2}{k_H^0}-\frac{\omega Q_2 \cos(\theta)}{k_H^0\sqrt{Q_2^2-m_b^{*2}}}\bigg( s-\frac{(\eta e)^2 a^2}{2(k\cdot p_1)}  \bigg).
\end{equation}
The term $|\overline{\mathcal{M}^{s}_{fi}}|^{2}$ can be calculated as follows:
\begin{equation}
|\overline{\mathcal{M}^{s}_{fi}}|^{2}=\frac{1}{2}\text{Tr}\Big[(\slashed{p}_2+m_b)\Big(C_0B_{s}(z)+C_1B_{1s}(z)+C_2B_{2s}(z)\Big)(\slashed{p}_1+m_t)\Big(\overline{C}_0B_{s}^*(z)+\overline{C}_1B_{1s}^*(z)+\overline{C}_2B_{2s}^*(z)\Big)\Big],
\end{equation}
with
\begin{equation}
\begin{split}
\overline{C}_0=&\mathcal{A}-\mathcal{B}\gamma_5,\\
\overline{C}_1=&C_{p_1}\slashed{a}_1 \slashed{k}(\mathcal{A}-\mathcal{B}\gamma_5)+C_{p_2}(\mathcal{A}-\mathcal{B}\gamma_5)\slashed{k}\slashed{a}_1,\\
\overline{C}_2=&C_{p_1}\slashed{a}_2\slashed{k}(\mathcal{A}-\mathcal{B}\gamma_5)+C_{p_2}(\mathcal{A}-\mathcal{B}\gamma_5)\slashed{k}\slashed{a}_2.
\end{split}
\end{equation}
\section{Results and discussion} \label{sect.2}
In this section, we present and analyze the numerical results obtained when the decay process occurs under the influence of a laser field. Prior to discussing these results, we note that the mass of the light charged Higgs boson, $M_{H^+}$, is varied in the range from 80 to 150 GeV, as LEP searches \cite{lep1,lep2,lep3} have excluded masses below 80 GeV. The ratio of the vacuum expectation values of the two Higgs doublets, $\tan \beta$, is set to 3, which represents a typical value within the viable parameter space of the type-I 2HDM. This choice is compatible with constraints from flavor physics, which disallow $\tan \beta \leq 2$ \cite{tan1,tan2}. We emphasize that our analysis is performed within the type-I 2HDM and focuses on a light charged Higgs boson, with $M_{H^+} \le m_t - m_b$, as required by the kinematics of the decay $t \to b H^+$. For heavier charged Higgs masses, this decay channel would be forbidden by energy conservation. The considered mass range is therefore physically motivated and consistent with the scope of the present study. In addition to these theoretical considerations, direct experimental searches at the LHC explicitly probe the light charged Higgs mass region relevant for $t \to b H^+$ decays. In particular, the ATLAS Collaboration has performed a search for a charged Higgs boson in the mass range 60--168~GeV in $pp$ collisions at $\sqrt{s}=13$~TeV with an integrated luminosity of 140~fb$^{-1}$, setting upper limits on the branching ratio $\mathrm{Br}(t \to bH^+)$ assuming $H^+ \to c s$ \cite{LHC}. Similar searches have also been performed by the CMS Collaboration in the same decay channel, setting upper limits on $\mathrm{Br}(t \to bH^+)$ for light charged Higgs masses 80--160~GeV in $pp$ collisions at $\sqrt{s}=13$~TeV \cite{CMS}. This demonstrates that the light charged Higgs scenario considered in this work is directly constrained and tested by current LHC experiments.\\
\begin{figure}[h!]
\centering
\includegraphics[scale=0.63]{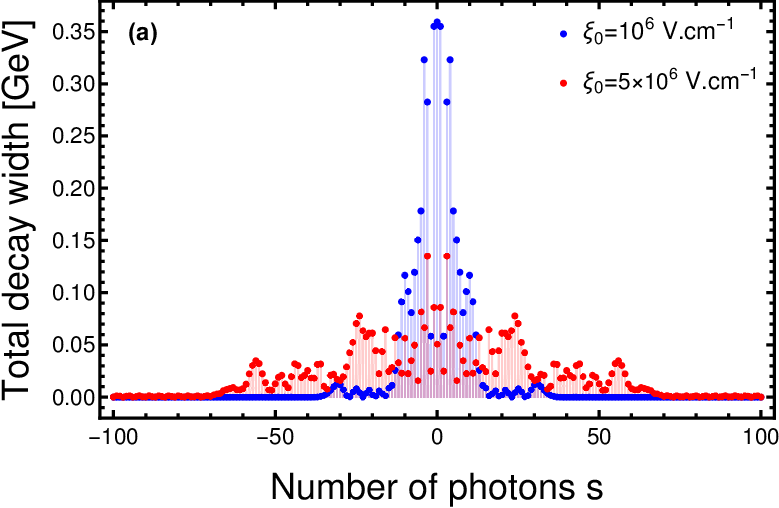} \vspace*{3mm}
\includegraphics[scale=0.554]{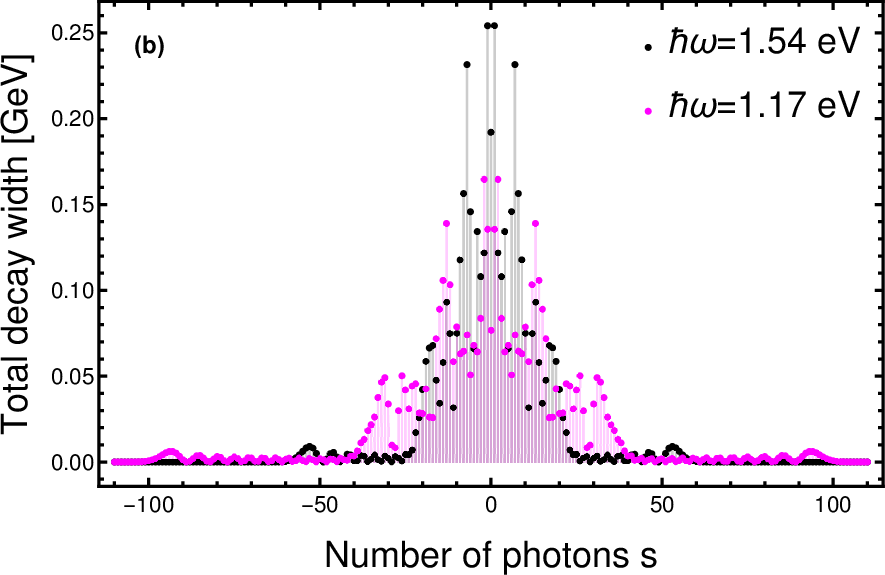} \\ \vspace*{2mm}
\includegraphics[scale=0.6]{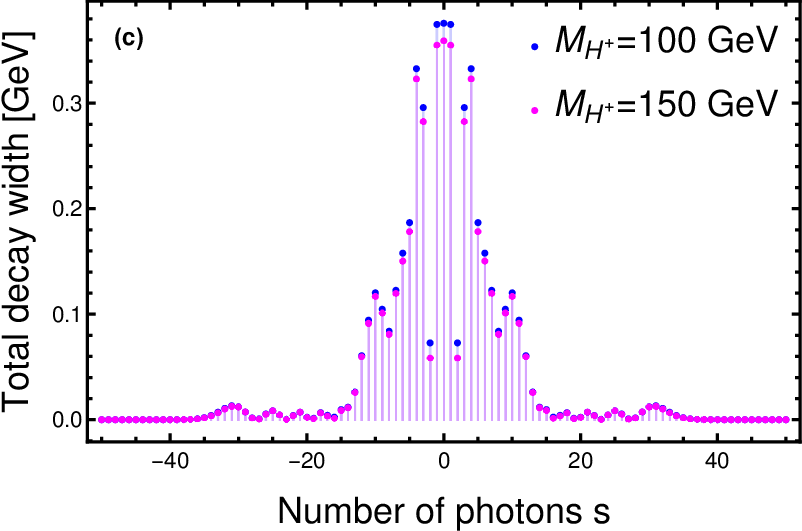} \vspace*{3mm}
\includegraphics[scale=0.6]{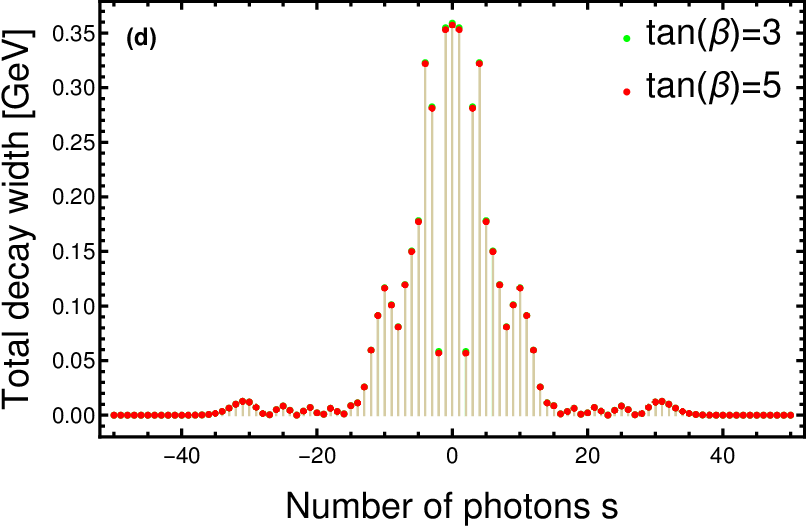} 
\caption{Dependence of the total decay width of the top quark on the number of absorbed photons $s$ in the presence of the laser field.  Unless stated otherwise, the parameters are: $M_{H^+} = 150$ GeV, $\tan \beta = 3$, $\xi_0 = 10^6$ V/cm, and $\hbar \omega = 2$ eV.}\label{fig2}
\end{figure}
Regarding the laser field parameters, we have considered values of field strengths and frequencies that are accessible with current laboratory setups and have been realized experimentally. The maximum field strength used remains below the Schwinger limit of $1.3\times 10^{16}$ V/cm, above which spontaneous electron-positron pair production occurs \cite{schwinger1,schwinger2,schwinger3}. The laser frequencies chosen correspond to commonly available laser sources, such as CO$_{2}$ lasers ($\hbar\omega = 0.117$ eV), Nd:YAG lasers ($\hbar\omega = 1.17$ eV), and Ti:sapphire lasers ($\hbar\omega = 1.54$ eV), ensuring that the scenario studied here is within the reach of current experimental technology. \\
We begin by examining changes in the total decay width of the top quark in the presence of the laser field in terms of the number of photons exchanged $s$. We get a set of envelopes as shown in Fig.~\ref{fig2}. Here, $s$ denotes the number of laser photons exchanged with the external field, with $s>0$ corresponding to photon absorption and $s<0$ to photon emission. The envelopes are symmetrical about $s=0$, indicating that the emission and absorption processes contribute equally to the total decay width. The total width drops to zero for large $|s|$ due to the cutoff number. Physically, this occurs because absorbing or emitting a very large number of photons would require more energy than is available in the decay, making the process kinematically forbidden. Mathematically, this behavior arises from the properties of the Bessel functions $J_s(z)$, which decrease rapidly when the order $s$ exceeds their argument $z$ \cite{mott,mott1}, reflecting the diminishing probability of such multi-photon processes. Both absorption and emission channels therefore contribute to the total decay width, while processes beyond the cutoff are effectively suppressed.
\\
Figure~\ref{fig2}(a) shows how the photon exchange varies as the field strength increases and the other variables are fixed at the following values: $M_{H^+}=150$ GeV, $\tan \beta=3$ and $\hbar\omega=2$ eV. We see that at the field strength $\xi_0=5\times 10^6$ V/cm, a large number of photons were exchanged compared to those exchanged at $\xi_0=10^6$ V/cm, demonstrating the strong interaction between the laser field and the decay system at high laser field strengths. The higher the laser field strength, the more pronounced is its effect on the decay width.  In Fig.~\ref{fig2}(b), we show the effect of the laser field frequency on the photon exchange between the laser field and the decaying system. In contrast to what we observed in Fig.~\ref{fig2}(a), the photon exchange appears to be significant at low frequencies (e.g., here the frequency is 1.17 eV), and decreases as the laser frequency increases. We will now examine the effect of the model parameters, $\tan \beta$ and the mass of the charged Higgs $M_{H^+}$, on the photon exchange process. From Figs.~\ref{fig2}(c) and \ref{fig2}(d), we observe that variations in these parameters produce almost identical envelopes, indicating that their effect on the number of photons exchanged is negligible within our present approximation. This is because the charged Higgs boson is not dressed by the laser field in the final state, and thus the influence of $\tan \beta$ and $M_{H^+}$ on the photon exchange process is expected to be minimal in this context. It should be emphasized that Fig.~\ref{fig2}(a) corresponds to a low-field (perturbative) regime and is included for illustrative purposes only. In the following analysis, we consider higher field strengths, $\xi_0 \gtrsim 10^{10}$ V/cm, in order to approach and satisfy the strong-field condition discussed at the end of this section. \\
\begin{table}[h]
\caption{Branching ratios of the top quark decay as a function of the laser field strength. 
The parameters are taken as $\hbar \omega = 0.117$ eV, $M_{H^+} = 150$ GeV, and $\tan \beta = 3$.}\label{tab1}
\begin{center}
\begin{tabular}{ccc}
\toprule
$\xi_0 [\text{V/cm}]$ & BR$(t\rightarrow bH^{+})$  & BR$(t\rightarrow bW^{+})$  \\ \hline
 $10^{10}$   & $0.009120$  &  $0.990880$  \\
 $10^{11}$   & $0.009190$  &  $0.990810$  \\
 $10^{12}$   & $0.009342$  &  $0.990658$  \\
 $2\times10^{14}$   & $0.046507$  &  $0.953493$  \\
 $3\times10^{14}$   & $0.795440$  &  $0.204560$  \\
 $3.8\times10^{14}$   & $0.975017$  &  $0.024983$  \\
 $4\times10^{14}$   & $0.920270$  &  $0.079723$  \\
 $5\times10^{14}$   & $0.365705$  &  $0.634295$  \\
 $6\times10^{14}$   & $0.157829$  &  $0.842171$  \\
 $7\times10^{14}$   & $0.115599$  &  $0.884401$  \\
 \hline
\end{tabular}
\end{center}
\end{table}\\
In Table~\ref{tab1}, we give the numerical values of branching ratios for top decay as a function of the laser field strength, with $\hbar\omega=0.117$ eV and $M_{H^+}=150$ GeV. We note that only field strengths satisfying $\xi_0 \gtrsim 10^{10}$ V/cm are retained, ensuring consistency with the strong-field condition discussed above. From this table, we can see that in the range of field strengths $[10^{10} - 2\times10^{14}~\text{V/cm}]$, the laser field has no noticeable effect on the branching ratios, as the branching ratio of top decay to the boson $W^+$ remains completely dominant as in the absence of the laser field. However, once we reach $\xi_0=3\times10^{14}$ V/cm until $\xi_0=4\times10^{14}$ V/cm threshold, we observe a significant change in the values of the branching ratios, where the branching ratio for the top decay to charged Higgs $H^+$ becomes dominant, reaching a value of BR$(t\rightarrow bH^{+})=0.975017$ at the field strength $\xi_0=3.8\times10^{14}$ V/cm. After that, when the field strength is above $4\times10^{14}$ V/cm, the branching ratio BR$(t\rightarrow bH^{+})$ continues to decrease. Note that the two branching ratios are always complementary, as their sum is necessarily equal to 1. Therefore, it is clear that these ratios are sensitive to the values of the chosen field strengths, as the branching ratio BR$(t\rightarrow bH^{+})$ exhibits complete dominance in the range of field strengths $[3\times10^{14} - 4\times10^{14}~\text{V/cm}]$.  Given this, whenever we want to fix the laser field strength, we take the value $\xi_0=3.8\times10^{14}$ V/cm, which gives a maximum value for the branching ratio of the charged Higgs boson. The rapid enhancement of BR$(t \to bH^+)$ for $\xi_0 \sim 3-4\times10^{14}$ V/cm originates from the photon-assisted nature of the decay in a strong laser field. At lower field strengths, the argument $z\propto \xi_0/\omega^2$ of the Bessel functions in the decay amplitude is small, leading to negligible multi-photon contributions and a suppressed BR$(t \to bH^+)$. Once the field strength reaches the threshold region, multi-photon processes become significant, causing the branching ratio to rise steeply. Beyond this range, the oscillatory behavior of the Bessel functions results in a decrease of BR$(t \to bH^+)$. Although the exact threshold depends on the chosen parameters ($M_{H^+}$, $\tan\beta$, and $\hbar \omega$), this qualitative behavior is a general feature of laser-assisted decays; see for example \cite{wdecay,zdecay}.\\
\begin{table}[h]
\caption{Branching ratios of the top quark decay as a function of the charged Higgs mass. 
The parameters are set to $\xi_0 = 3.8\times10^{14}$ V/cm,  $\hbar \omega = 0.117$ eV, and $\tan \beta = 3$.}\label{tab2}
\begin{center}
\begin{tabular}{ccc}
\toprule
$M_{H^{+}}[\text{GeV}]$ & BR$(t\rightarrow bH^{+})$  & BR$(t\rightarrow bW^{+})$  \\ \hline
 $80$   & $0.984625  $       &  $ 0.0153749$  \\
 $90$   & $ 0.983693$      &  $ 0.0163071$  \\
 $100$   & $ 0.982434$     &  $ 0.017566$   \\
 $110$   & $ 0.980777$     &  $ 0.0192231$   \\
 $120$   & $ 0.978866$      &  $0.021134 $   \\
 $130$   & $ 0.976474$      &  $ 0.0235256$   \\
 $140$   & $0.973417 $      & $ 0.0265825$   \\
 $150$   & $ 0.975017$      &  $ 0.0249832$   \\
 \hline
\end{tabular}
\end{center}
\end{table}
To explore the dependence on the charged Higgs boson mass, Table~\ref{tab2} presents the numerical values of the branching ratios for top decay as a function of the charged Higgs mass, assuming $\xi_0 = 3.8 \times 10^{14}$ V/cm and $\hbar\omega = 0.117$ eV. Table~\ref{tab2} shows that for a laser field strength of $3.8 \times 10^{14}$ V/cm and a frequency of $0.117$ eV, the branching ratio BR$(t\rightarrow bH^{+})$ reaches its maximum and becomes dominant for all charged Higgs masses between 80 and 150 GeV. This means that the laser field used here with carefully chosen field strength and frequency values can only produce a charged Higgs boson $H^+$ with a mass $M_{H^+}$ between 80 and 150 GeV. Otherwise, only the branching ratio of decay to the $W^+$ boson is possible. We note that the numerical values presented in Table~\ref{tab2} are obtained by truncating the photon-number summation over a finite range of $s$. A detailed convergence analysis shows that, near the upper end of the considered mass range ($M_{H^+}=140$--150 GeV), a limited truncation may induce small numerical deviations. By extending the summation up to $|s|\leq 2000$, the branching ratios become fully stable and converge to nearly identical values, with differences below $10^{-5}$. This confirms that the branching ratio $\mathrm{BR}(t\to bH^+)$ decreases monotonically with increasing $M_{H^+}$, as expected from phase-space suppression, and that the slight enhancement observed at $M_{H^+}=150$ GeV in Table~\ref{tab2} is a numerical artifact rather than a physical effect.\\
\begin{figure}[hbtp]
\centering
\includegraphics[scale=0.65]{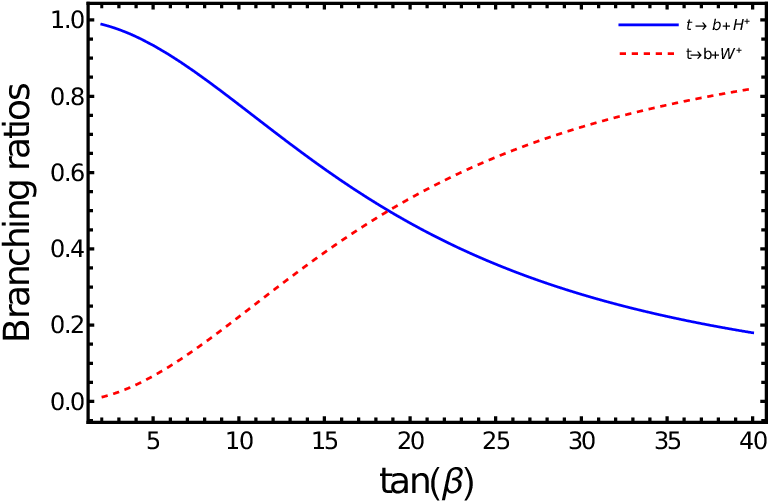}
\caption{Behavior of branching ratios for top decay as a function of $\tan \beta$, with $\xi_0=3.8\times10^{14}$ V/cm, $\hbar\omega=0.117$ eV and $M_{H^+}=150$ GeV.}\label{fig3}
\end{figure}
Now, we turn to the dependence on $\tan \beta$. To analyze the impact of $\tan \beta$, we examine how the branching ratios vary with different values of $\tan \beta$, keeping other parameters fixed. Figure~\ref{fig3} presents the behavior of branching ratios for top decay as a function of $\tan \beta$, with $\xi_0=3.8\times10^{14}$ V/cm, $\hbar\omega=0.117$ eV and $M_{H^+}=150$ GeV. From Fig.~\ref{fig3}, the parameter $\tan \beta$ influences the behavior of the branching ratios. The region of interest is where the charged Higgs branching ratio is dominant, which occurs for $ 3 \leq \tan \beta \leq 18 $, with a maximum at $\tan \beta= 3$. Beyond this range, the branching ratio BR$(t\rightarrow bH^{+})$ gradually decreases. This is expected because, in the type-I 2HDM, all the Yukawa couplings of $H^+$ are inversely proportional to $\tan \beta$. The benchmark choice $\tan\beta=3$ is adopted in figures and tables where $\tan\beta$ is fixed, since it lies in the allowed region of the type-I 2HDM and maximizes the branching ratio $\mathrm{BR}(t \to bH^+)$, while the qualitative enhancement persists over the wider range $2 \leq \tan\beta \leq 18$.\\
\begin{figure}[hbtp]
\centering
\includegraphics[scale=0.65]{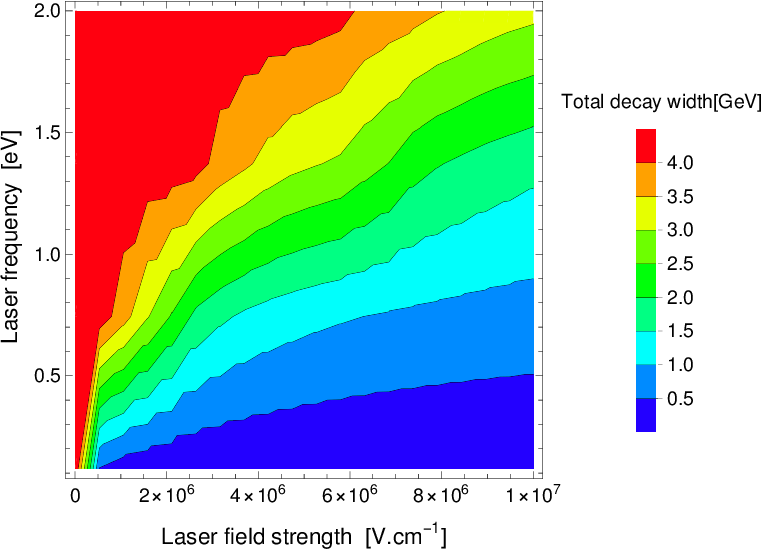}
\caption{Contour plot illustrating the total decay width of the top quark in the presence of a laser field, as a function of the field strength and frequency. The parameters are $M_{H^+} = 150$ GeV and $\tan \beta = 3$.}\label{fig4}
\end{figure}
To further investigate the impact of the laser field on the total decay width, we present in Fig.~\ref{fig4} a contour plot illustrating its variations as a function of both the laser field strength and frequency. This representation allows us to analyze how these two parameters simultaneously influence the total decay width. The plot highlights regions where the decay width reaches its maximum and reveals the parameter space where significant variations occur. From Fig.~\ref{fig4}, we observe that the total decay width varies with laser field strength and frequency. As the frequency increases, the laser field's influence on the decay width diminishes. However, at each fixed frequency, the effect becomes more significant with increasing field strength. Consequently, the most pronounced effect occurs at high field strengths and low frequencies, in agreement with the photon exchange process discussed in Figs.~\ref{fig2}(a) and \ref{fig2}(b).

To ensure the consistency of our results, we conclude our discussion by assessing the validity of the Dirac-Volkov formalism in the parameter regimes considered. The validity of the Dirac-Volkov formalism is governed by the dimensionless intensity parameter 
\begin{equation}
a_0 = \frac{|e|\xi_0}{m \omega},
\end{equation}
which characterizes the strength of the interaction between charged particles and the external electromagnetic field. The strong-field (Volkov) regime corresponds to $a_0 \gtrsim 1$, while for $a_0 \ll 1$ the interaction remains perturbative.
For the parameters used in Fig.~\ref{fig2}(a), namely $\xi_0 = 10^6$--$5\times10^6$ V/cm and $\hbar\omega = 2$ eV, one obtains $a_0 \sim 10^{-5}$, indicating that this regime is not strictly in the strong-field domain. This figure is therefore intended to provide a qualitative illustration of the dependence of photon exchange on the field strength. In contrast, for the physically relevant regime considered in Table~\ref{tab1}, with $\hbar\omega = 0.117$ eV and $\xi_0 \sim 3\times10^{14}$--$4\times10^{14}$ V/cm, the parameter $a_0$ reaches larger values, satisfying the strong-field condition $a_0 \gtrsim 1$. This confirms the validity of the Dirac-Volkov formalism in the regime where significant modifications of the decay dynamics are observed.
\section{Conclusion} \label{sect.3}
In this study, we analyzed the influence of a circularly polarized laser field on the top quark decay into a charged Higgs boson within the type-I 2HDM. Using the Dirac-Volkov formalism, we demonstrated that an external electromagnetic field can significantly modify the decay dynamics, affecting the branching ratios of different channels. Our results show that, at a laser field strength of $3.8\times 10^{14}$ V/cm and a frequency of $0.117$ eV, the branching ratio for the decay into a charged Higgs boson of mass between 80 and 150 GeV and a bottom quark reaches 0.97, making it the dominant channel over the standard  decay. This suggests that intense electromagnetic fields could enhance the possibility of detecting charged Higgs bosons in high-energy collisions, providing a new avenue for experimental verification of physics beyond the SM. These findings highlight the potential role of strong laser fields in collider physics, suggesting that future experiments could take advantage of laser-assisted processes to probe new physics scenarios. It is worth noting that the laser field strength of $3.8\times 10^{14}$ V/cm has not yet been achieved in current laboratory setups. However, with the rapid progress in laser technology, researchers anticipate that even higher intensities will become available in the near future, opening new possibilities for exploring strong-field effects in high-energy physics. We note that our calculation is performed at tree level. One-loop QCD corrections are known to modify the decay width by approximately $6-15\%$ \cite{loop1,loop2}, while electroweak loop contributions are typically smaller \cite{weak}. Since our main focus is on the laser-assisted modifications of the decay, these higher-order effects may shift the numerical values but are not expected to qualitatively change the overall behavior or main conclusions of our study.

\section*{Acknowledgments}  
We would like to thank M. Krab for valuable discussions and insightful comments on the variation of the parameter space in the type-I 2HDM.  


\end{document}